\documentstyle[preprint,aps,eqsecnum]{revtex}

\begin{document}
\draft
\preprint{gr-qc/9611064}

\title{A REVISED PROOF OF\\
THE EXISTENCE OF NEGATIVE ENERGY DENSITY\\ 
IN\\ 
QUANTUM FIELD THEORY}
\author{Chung-I Kuo\\
Department of Physics\\
Soochow University\\
Taipei, Taiwan, Republic of China}
\vskip 1 true cm
\maketitle

\begin{abstract}
Negative energy density is unavoidable in the quantum
theory of field. We give a revised proof of the 
existence of negative energy density unambiguously
for a massless scalar field.
\end{abstract}
\vskip 2 true cm
\pacs{03.65.Sq, 03.70.+k, 05.40.+j}
\vfill
\eject

\section{Introduction}

The notion of a state with negative energy is not 
familiar in the realm of classical physics.
However, it is not rare in quantum field theory
to have states with negative energy density.
Even for a scalar field in the
flat Minkowski spacetime, it can be proved that
the existence of quantum states with negative
energy density is inevitable \cite{Epstein65}. 

Although all known forms of classical matter 
have non-negative energy density, it is not so 
in quantum field theory. A general state can be a 
superposition of number eigenstates 
and may have a negative expectation value
of energy density in certain spacetime regions
due to coherence effects, thus 
violating the weak energy condition
\cite{Epstein65}. If there were no constraints
on the extent of the  violation of the weak energy
condition, several dramatic and disturbing
effects could arise. These include the
breakdown of the second law of thermodynamics 
\cite{Ford78}, of cosmic censorship \cite{Ford90}
, and of causality \cite{Morris88a} (for a more thorough
review and tutorial see \cite{Morris88b}). There are, 
however, two possible reasons as to why these effects will 
not actually arise. The first is the existence  
of constraints on the magnitude and the spatial 
or temporal extent of the negative energy 
\cite{Ford78,Ford90}. The second is 
that the semiclassical theory of gravity may 
not be applicable to systems in which the 
energy density is negative \cite{Kuo93,Kuo94}.

\section{Negative Energy Densities}
\label{sec:negative}
  
The quantum coherence effects which produce
negative energy densities can be easily 
illustrated by the state composed of two 
particle number eigenstates
\begin{equation}
|\Psi\rangle\equiv
{1\over\sqrt{1+\epsilon^2}
}\,\left(|0\rangle + \epsilon|2\rangle\right), 
\label{0+2}
\end{equation}
where $|0\rangle$ is the vacuum state 
satisfying $a|0\rangle=0$, 
and $|2\rangle=
{1\over\sqrt 2}(a^{\dagger})^2|0\rangle$ 
is the two particle state. Here we take $\epsilon$, the relative 
amplitude of the two states, to be real for simplicity. For this 
state \cite{Kuo93},
\begin{eqnarray}
\langle\colon T_{\alpha\beta}(x)\colon\rangle
&=&\langle \Psi|\colon T_{\alpha\beta}(x)\colon
|\Psi\rangle \nonumber \\
&=&{\epsilon\over 1+\epsilon^2}\,\lbrace{\sqrt 2}
(T_{\alpha\beta}[f_{\bf k},
f_{\bf k}]+T_{\alpha\beta}[f_{\bf k}^{\ast},
f_{\bf k}^{\ast}])+2\epsilon(T_{\alpha\beta}[f_{\bf k},
f_{\bf k}^{\ast}]+T_{\alpha\beta}[f_{\bf k}
^{\ast},f_{\bf k}])\rbrace \nonumber \\
&=&2\,{{{\cal K_{\alpha\beta}}\epsilon}\over{1+\epsilon^2}} 
\left(2\epsilon-{\sqrt 2} \cos\left(2\theta\right)\right).
\label{EnergyDensityOf0+2}
\end{eqnarray}
Obviously the energy density can be positive or negative depending 
on the value of $\epsilon$ and the 
spacetime-dependent phase $\theta\equiv 
k_{\rho}x^{\rho}$.   
We also observe that the negative
contribution comes from the cross term. For a
general state which is a linear combination
of $N$ particle number eigenstates, the number
of cross terms will increase as $N^2$.
Therefore, for a general quantum state, the
occurrence of a negative energy density is very
probable. 

Epstein et al \cite{Epstein65} have demonstrated 
the nonpositivity of the energy density of a free massive 
scalar field theory in axiomatic quantum field theory. 
The above illustrative example has already given us 
some physical intuition about it. Here we are 
going to present a re-derivation of parts of the original 
proof, which is somewhat vague in several points.
We will restrict our concern to the massless case
so that we can easily follow the notations used 
in \cite{Kuo93,Kuo94}, which will not change the conclusion
anyway.

A field at a spacetime point is meaningless in 
quantum theory.
It must be defined in a distributional sense.
So the corresponding question of the positivity of 
the energy density in quantum field theory would 
be that if for all $|\Phi\rangle$
\begin{equation}
\langle \Phi|\,\colon T_{00}(f)\colon\,|\Phi\rangle\geq 0
\end{equation}
or not. Notice that here we have the same form of 
stress tensor as discussed before and
\begin{equation}
T_{00}(f)=\int\,dx\, T_{00}(x)\,f(x)
\end{equation}
is a distribution-valued operator with respect to 
the test function $f(x)$. Here $x$ can be any 
spacetime variables. For example,
\begin{equation}
T_{00}(f(t))=\int\,dt\,{T_{00}(t)\over{t^2+t_0^2}}
\end{equation}
represents the measurement of the energy density
is performed in a finite range of time $t_0$.
The answer is surely ``no" as we can see from the
above simple example illustrating the quantum
coherence effect.

Suppose we choose a positive-definite
test function $f(x)$. For simplicity we can also 
switch our discussion to a local field $T$ 
instead of $T_{00}$.
We can decompose a general quantum state (not the
vacuum state in particular) 
$|\Phi\rangle$ into superposition of particle 
number eigenstates as
\begin{equation}
|\Phi\rangle=\sum_{r=0}^{n}\,c_r\,|r\rangle ,
\end{equation}
where $|r\rangle$ are the $r$ particle states.
Because of the bi-linearity of the form of the 
stress tensor of a free scalar field, we 
can decompose the normal-ordered stress tensor
$\colon T \colon$ into
\begin{equation}
\colon T \colon = T^{(1)}\,(a^{\dagger})^2
+T^{(2)}\,a^{\dagger}\,a +T^{(3)}\,a^2 .
\end{equation}
Accordingly, the $m$-th power 
of normal-ordered \,$T$\, is
\begin{eqnarray}
\colon T \colon ^m &=&
\colon\left( T^{(1)}\,(a^{\dagger})^2
+T^{(2)}\,a^{\dagger}\,a +
T^{(3)}\,a^2\right)\colon ^m \nonumber\\ 
&=& \sum_{i,j=0}^m\,{m!\over{i!\,(j-i)!\,(m-j)!}}\,
(T^{(1)})^i\,(T^{(2)})^{j-i}\,(T^{(3)})^{m-j}\,
(a^{\dagger})^{i+j}\,a^{2m-i-j},\nonumber\\
\end{eqnarray}
in which we have neglect the ordering of $a$
and $a^{\dagger}$ since not the ordering but the 
power of them concerns us here.

As long as the test function $f$ is positive 
definite and will not 
affect our discussion of the sign, we can leave 
it out and then insert it back later when needed. 
The expectation value is
\begin{eqnarray}
\langle \Phi|\,\colon T\colon ^m |\Phi\rangle &=&
\sum_{i,j=0}^m\sum_{r,s=0}^n\,c_r^\ast\,c_s
\,{m!\over{i!\,(j-i)!\,(m-j)!}}\,\times\nonumber\\
&&(T^{(1)})^i\,(T^{(2)})^{j-i}\,(T^{(3)})^{m-j}\,
\langle r|\,(a^{\dagger})^{i+j}\,a^{2m-i-j}\,
|s\rangle \nonumber\\
&=&
\sum_{i,j=0}^m\sum_{r,s=0}^n\,c_r^\ast\,c_s\,
{m!\over{i!\,(j-i)!\,(m-j)!}}\,(r!)^{-1/2}(s!)^{-1/2}
\times\nonumber\\
&&(T^{(1)})^i\,(T^{(2)})^{j-i}\,(T^{(3)})^{m-j}\,
\langle 0|\,a^r\,(a^{\dagger})^{i+j}\,a^{2m-i-j}\,
(a^{\dagger})^s\,|0\rangle .\nonumber\\
\end{eqnarray}
Since the highest value $r$ or $s$ can take is
$n$, if we choose $m>n$, either the right (when 
$i+j< n$) or the left operation (when 
$i+j> n$) will give us a vanishing result. 
So, for large enough $m$, we can always make
$\langle \Phi|\,\colon T\colon ^m |\Phi\rangle$ 
vanishing. This is an important point missing in the 
original proof of \cite{Epstein65}.
From now on we will omit the normal
ordering notation and implicitly assume it for
simplicity.

For arbitrary states $|\Psi\rangle$ and 
$|\Phi\rangle$, since 
$\langle \Phi|\,T\,|\Phi\rangle$ and
$\langle \Psi|\,T\,|\Psi\rangle$ are both 
non-negative as assumed, we may use the Schwartz 
inequality for the state vectors in the Hilbert 
space
\begin{equation}
|\langle \Psi|\,T\,|\Phi\rangle |^2 \leq 
\langle \Phi|\,T\,|\Phi\rangle\,
\langle \Psi|\,T\,|\Psi\rangle .
\label{Schwartz}
\end{equation}
That means $\langle \Phi|\,T\,|\Phi\rangle=0$ implies
$\langle \Psi|\,T\,|\Phi\rangle=0$. It follows that
$T\,|\Phi\rangle=0$ since $| \Psi\rangle$ is 
arbitrary. The inequality is surely true when $T$
is the identity operator, in which case the
inequality just states that the cosine theorem 
holds for the state vectors in the Hilbert space.
It is essential to note in the proof that for
a general operator $T$, we must assume the 
positive definiteness of the expectation values
of $T$ for general quantum states to have the
inequality hold. We will utilize this inequality
with other facts to deduce a contradiction and 
thus invalidate the positive definiteness of the 
expectation values of $T$.

If $m$ is an even integer, we have
\begin{equation}
\langle \Phi|\,T^m\,|\Phi\rangle=
\langle T^{m/2}\,\Phi\,|\,T^{m/2}\,\Phi\rangle=0,
\end{equation}
from which follows $T^{m/2}\,|\Phi\rangle=0$.
If $m$ is an odd integer, we then have
\begin{equation}
\langle \Phi|\,T^m\,|\Phi\rangle=
\langle T^{[m/2]}\,\Phi\,|\,T\,|\,T^{[m/2]}\,\Phi\rangle=0,
\end{equation}
which follows $T^{[m/2]+1}\,|\Phi\rangle=0$
by the argument derived from Schwartz inequality. 
We can take the inner product of  
$T^{[m/2]}\,|\Phi\rangle$ or
$T^{[m/2]+1}\,|\Phi\rangle$ with $\langle\Phi |$
and repeat the same procedure until we get 
$T\,|\Phi\rangle=0$.

\section{Conclusion}

In conclusion, we can always find a large enough 
integer $m$ to make 
$\langle\Phi|\,T(f)^m\,|\Phi\rangle=0$ ($f(x)$ 
inserted back will not affect all the previous 
analysis). However, by dividing $T^m$ into halves 
and applying in both directions repeatedly, it 
implies $T\,|\Phi\rangle=0$
for arbitrary $|\Phi\rangle$, which in turn implies
$T\equiv 0$. We start with a general $T$, and 
end up with the trivial case $T\equiv 0$, which 
is obviously not what we are interested in. The 
contradiction denotes the original assumption of 
the non-negativeness of energy density used in 
Schwartz inequality (\ref{Schwartz})
\begin{equation}
\langle \Phi|\,T_{00}(f)\,|\Phi\rangle\geq 0
\end{equation}
is incorrect. Therefore we arrive at the conclusion
that in quantum field theory it is possible for 
general quantum states to have negative energy 
densities. The same argument about the positive 
non-definiteness also apples to any component of
the stress tensor.

\section*{Acknowledgement}
The author would like to thank Professor L. H. Ford, 
and Professor T. Roman for reading the manuscript.

\end{document}